\documentclass[traditabstract,times]{aa} 

\newcommand{\ergps}{erg\thinspace s$^{-1}$}
\newcommand{\ergpspsqcm}{erg\thinspace s$^{-1}$\thinspace cm$^{-2}$}
\newcommand{\psqcm}{cm$^{-2}$}
\newcommand{\nH}{$N_{\rm H}$}

\usepackage{graphicx}
\begin{document}
\title{The location of an active nucleus and the soft X-ray shadowing by a tidal tail in the ULIRG Mrk 273}


   \author{K. Iwasawa\inst{1}\thanks{Email: kazushi.iwasawa@icc.ub.edu}
          \and
          J.M. Mazzarella\inst{2}
          \and
J.A. Surace\inst{3}
\and
D.B. Sanders\inst{4}
\and
L. Armus\inst{3}
\and
A.S. Evans\inst{5}
\and
J.H. Howell\inst{3}
\and
S.~Komossa\inst{6}
\and
A. Petric\inst{3}
\and
S.H. Teng\inst{7}
\and
Vivian U\inst{4,8}
\and
S. Veilleux\inst{7}
          }

          \institute{ICREA and Institut de Ci\`encies del Cosmos, Universitat de Barcelona, Mart\'i i Franqu\`es, 1, 08028 Barcelona, Spain
            \and
            IPAC, California Institute of Technology, Pasadena, CA 91125, USA
            \and
            Spitzer Science Center, California Institute of Technology, Pasadena, CA 91125, USA
\and
            Institute for Astronomy, 2680 Woodlawn Drive, Honolulu, HI 96822-1839, USA
            \and
            Department of Astronomy, University of Virginia, 530 McCormick Road, Charlottesville, VA 22904 and NRAO, 520 Edgemont Road, Charlottesville, VA 22903-2475, USA
\and
Max Planck Institut f\"ur extraterrestrische Physik, Gie\ss enbachstra\ss e, 85748 Garching, Germany
\and
Department of Astronomy, University of Maryland, College Park, MD 20742, USA
\and
Harvard-Smithsonian Center for Astrophysics, Cambridge, MA, 02138, USA
          }


 
   \abstract{ Analysis of data from the Chandra X-ray Observatory for
     the double nucleus ULIRG Mrk 273 reveals an absorbed hard X-ray
     source coincident with the southwest nucleus, implying that this
     unresolved near infrared source is where an active nucleus
     resides while the northern nuclear region contains a powerful
     starburst which dominates the far infrared luminosity. There is
     evidence of a slight image extension in the 6-7 keV band, where a
     Fe K line is present, towards the northern nucleus. A
     large-scale, diffuse emission nebula detected in soft X-rays
     contains a dark lane that spatially coincides with a high
     surface-brightness tidal tail extending $\sim 50$ arcsec (40 kpc)
     to the south. The soft X-ray source is likely located behind the
     tidal tail which absorbs X-ray photons along the line of
     sight. The estimated column density of cold gas in the tidal tail
     responsible for shadowing the soft X-rays is $N_{\rm H}\geq
     6\times 10^{21}$ cm $^{-2}$, consistent with the tidal tail having
     an edge-on orientation.}

\titlerunning{The nuclear position of Mrk 273 and the soft X-ray tail}
\authorrunning{K. Iwasawa et al.}
   \maketitle
%

\section{Introduction}

Mrk 273 (= UGC 8696, IRAS F13428+5608) is a nearby ($z=0.0378$)
ultra-luminous infrared galaxy (ULIRG) with a 8-1000 $\mu $m
luminosity of $1.6\times 10^{12}L_{\odot}$. The galaxy has a long
tidal tail extending $\sim 40$ kpc to the south, indicating that the
system is an ongoing merger. The nuclear region appears complex and
two nuclei with a projected separation of $\sim 1^{\prime\prime}$ have
been revealed by infrared imaging (Majewski et al 1993; Knapen et al
1997; Scoville et al 2000). In low resolution radio images, three
radio components are present (Condon et al 1991), and the northern (N)
and southwest (SW) components are identified with the two infrared
nuclei while the southeast (SE) component is likely a star cluster
seen in the Hubble Space Telescope (HST hereafter) ACS and NICMOS
images (Scoville et al 2000). Mrk 273 N is a much stronger radio
source with resolved structures in the high-resolution imaging (Knapen
et al 1997; Carilli \& Taylor 2000; Bondi et al 2005). It is the site
of a large concentration of molecular gas containing $1\times 10^9
M_{\odot}$ in a disk traced by CO(2-1) (Downes \& Solomon
1998). Combined with the radio continuum morphology which indicates an
abundance of supernovae and supernova remnants (Carilli \& Taylor
2000; Bondi et al 2005), this region is likely to contain a powerful
starburst. The 1-17 $\mu $m spectral energy distribution (SED) of the
N and SW nuclei obtained by high resolution imaging with the MIRLIN
mid-infrared camera on Keck (Soifer et al 2000) suggests that strong
PAH features detected in the ISO SWS (Genzel et al 1998) and Spitzer
IRS (Armus et al 2007) spectra originate from the N nucleus, in
agreement with a strong starburst.

The presence of an active nucleus (AGN) has been indicated by the
optical spectrum (Khachikian \& Weedman 1974; Sanders et al 1988),
which is supported by the detection of [Ne {\sc v}]$\lambda
14.3\thinspace\mu $m and [O{\sc iv}]$\lambda 25.9\thinspace\mu$m
(Genzel et al 1998; Satyapal et al 2004; Armus et al 2007) and an
absorbed hard X-ray source with $N_{\rm H}\simeq 4\times 10^{23}$
\psqcm (Iwasawa 1999; Risaliti et al 2000; Balestra et al 2005), which
is variable (Teng et al 2009). Scoville et al (2000) suggested that
the SW nucleus hosts the AGN because it is red and unresolved in the
NICMOS K-band image. Although the SW component has only a very weak
extended radio continuum, its 1-17 $\mu $m SED (Soifer et al 2000)
shows that the SW nucleus is comparable or brighter at 3.4 and 10.3
$\mu $m than the N nucleus, suggesting a hot/warm dust
component. Integral field spectroscopy of the nuclear region by Colina
et al (1999) shows that Seyfert 2 type excitation with strong [O{\sc
  iii}] $\lambda 5007$ is found at the SW nucleus while the N nucleus
shows characteristics of a LINER (in agreement with the classification
of Veilleux et al 1995). However, a point-like hard X-ray source
imaged by the Chandra X-ray Observatory (hereafter Chandra) has been
identified with the N nucleus (Xia et al 2002; Gonz\'alez-Mart\'in et
al 2006). In this paper, we re-examine this identification, as Chandra
astrometry has been improved since the reprocessing of the data with
the new alignment file in 2007. A combination of a re-analysis of the
Chandra data and an accurate astrometry on the HST images enables us
to locate the hard X-ray source more accurately.

On a larger scale, there is soft X-ray emission extending $\sim 30$
kpc to the south. In the same region, a long optical tidal tail with a
similar length is seen. The optical surface brightness of this tidal
tail is one of the highest observed in a large sample of
interacting/merging galaxies (Mazzarella \& Boroson 1993), suggesting
it has a rare edge-on orientation in the line of sight. There should
be no direct physical connection between the optical and X-ray features however,
because one is the remnant of a galaxy merger made of stars and the
other is likely hot gas in a galactic-scale outflow (``superwind'',
e.g., Heckman, Armus \& Miley 1990).

Mrk 273 is in the C-GOALS sample for which general X-ray properties
obtained from Chandra observations are presented in Iwasawa et al
(2011). Further detailed study of the Chandra data are found in Xia et
al (2002, see also Ptak et al 2003; Grimes et al 2005; Satyapal et al
2004; Gonz\'alez-Mart\'in et al 2006). In this article, we focus on 1)
the hard X-ray imaging of the nuclear region; and 2) the large,
southern soft X-ray nebula and its relation to the tidal tail.

The cosmology adopted here is $H_0=70$ km s$^{-1}$ Mpc$^{-1}$,
$\Omega_{\Lambda}=0.72$, $\Omega_{\rm M}=0.28$, consistent with
that adopted for other objects in the Great Observatories All Sky LIRG
Survey (GOALS, Armus et al 2009). Thus the luminosity distance of
$D_{\rm L}=173$ Mpc and the angular scale of 0.77 kpc arcsec$^{-1}$
are assumed.

\section{Observation and data reduction}

Mrk 273 was observed with Chandra on 2000 April 19 (Obs ID 809) using
the focal plane detector ACIS-S operated in VFAINT imaging mode. The
nucleus of the galaxy was placed at the off-axis angle of 1 arcmin on
the detector. The useful exposure time is 44.2 ks. The data were
reduced with the standard Chandra data analysis package CIAO 4.2 with
the latest calibration files in CALDB 4.3.

We use the archival H band image of Mrk 273 obtained from the
HST-NICMOS with the NIC2 F160W filter (Scoville et
al 2000). The optical B and I band images were taken by the HST-ACS
with the F435W and F814W filters, respectively, as part of GOALS
(Evans et al in prep.).

The astrometry correction of the HST image was carried out with the
Starlink software GAIA. The X-ray spectral analysis was performed
using the HEASARC's FTOOLS and XSPEC version 12.

\section{Results}

\subsection{The position of the hard X-ray source}

\begin{figure*}
\centering
\centerline{\includegraphics[width=\textwidth,angle=0]{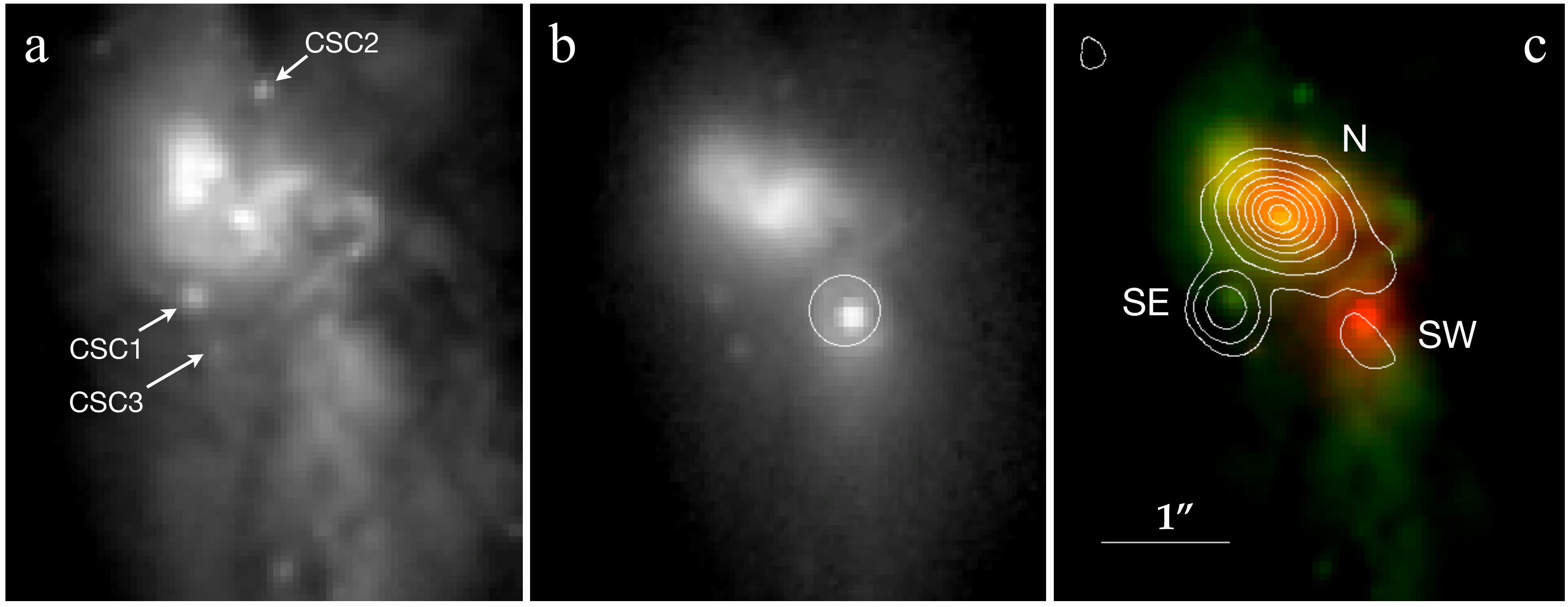}}
\centerline{\includegraphics[width=\textwidth,angle=0]{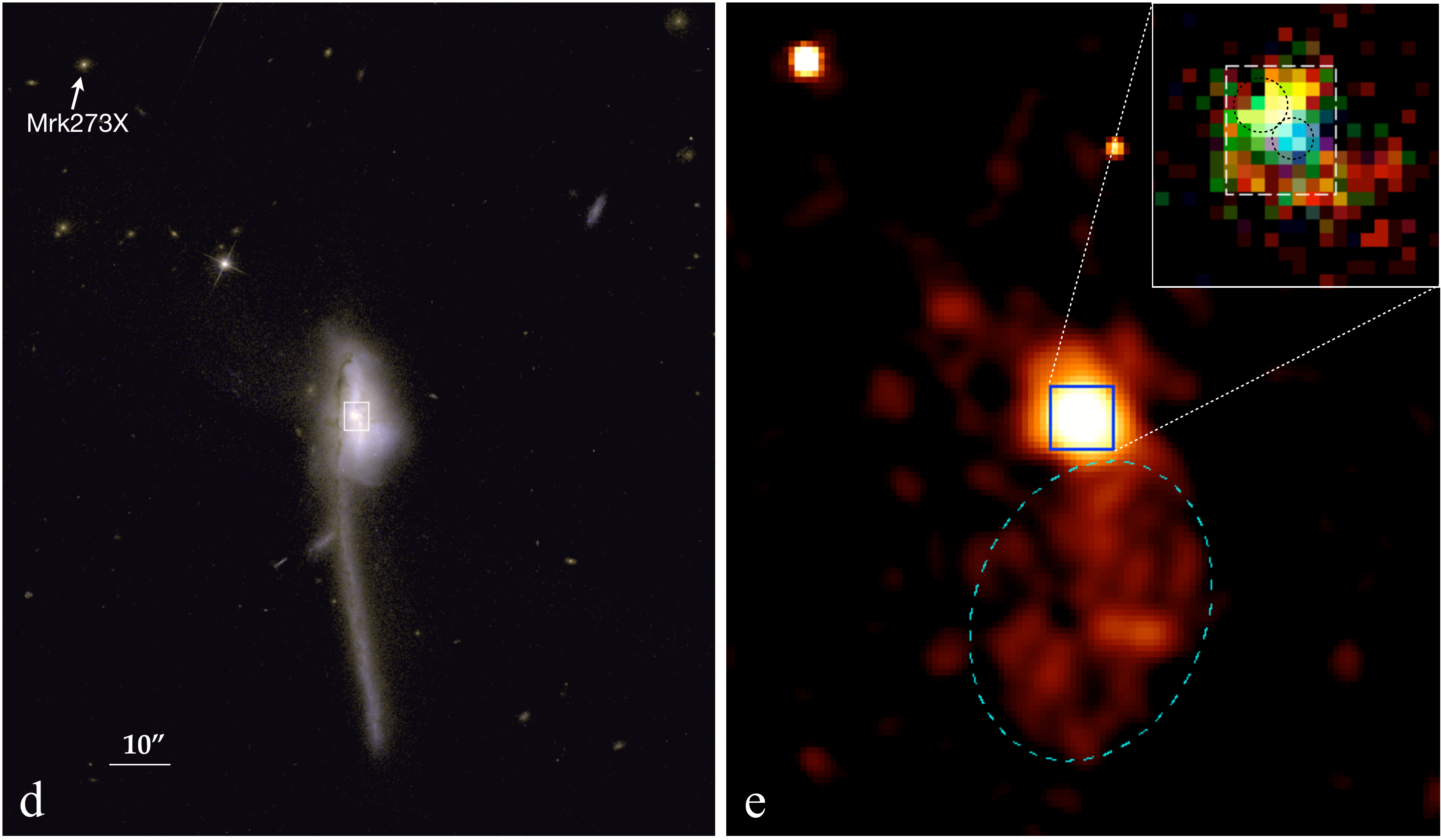}}
\caption{ a) The HST-ACS I band (F814W) image of the nuclear region of
  Mrk 273. Candidate star clusters used to register the NICMOS image
  with the ACS image are labeled in order of their luminosity. The
  orientation of this and all the other images is north up, east to
  the left. b) The HST-NICMOS H-band (F160W) image of the same region
  as in a). The point-like SW nucleus, which is not evident in the
  optical band, is clearly visible. The centroid of the hard X-ray
  (4-6 keV) source is indicated by the error circle of 0.35 arcsec
  radius. c) The composite of the I-band (green) and H-band (red)
  images of the same region as in a) and b), overlaid by the VLA 8.4
  GHz image (Condon et al 1991) in contours. No astrometric correction
  has been applied to the VLA image. The three radio components, N, SE
  and SW are labeled. The lowest contour represents 8\% level of the
  peak brightness at the N component and other contours increases by
  factors of 3.3. The scale bar indicates 1 arcsec ($=0.77$ kpc). d)
  The HST-ACS B (F435W)$+$I (F814W) composite image of a large field of view
  ($2.0^{\prime}\times 2.3^{\prime}$) around Mrk 273. The area of the
  images for the nuclear region in a), b) and c) is indicated by a
  white rectangle. The bright, point-like X-ray source Mrk 273 X, a
  distant ($z=0.46$) Seyfert 2 galaxy, seen in the X-ray image
  (Fig. 1e) and used to register the Chandra image with the ACS image
  is labeled. The scale bar indicates 10 arcsec ($=7.7$ kpc). e) The
  Chandra ACIS-S 0.4-7 keV image of the same area of the sky as in
  d). Details of the bright, central part are shown in the inset:
  Three colour composite of unsmoothed images of the 0.4-1.1 keV (red),
  1.1-3 keV (green) and 3-7 keV (blue) bands for the $10^{\prime\prime}\times
  10^{\prime\prime}$ region, as marked in the main figure. The two
  black, dashed circles indicate the regions where the spectral data
  (N and SW) shown in Fig. 3 were taken. The SW region is centred on
  the hard X-ray source, where the X-ray colour is the bluest. The
  white dashed rectangle indicates the region for a), b) and c). The
  ellipse in the main figure indicates the region where the spectral
  data for the southern X-ray nebula (Fig. 5) were collected.}
\end{figure*}

The compact hard X-ray source is located where the soft X-ray emission
is suppressed due to obscuration and thus the brightest soft X-ray
region is displaced to the north by $\sim 1$ arcsec (see Iwasawa et al
2011; Xia et al 2002). We restrict the energy range of the image to 4-6
keV band, where the image is close to point-like. 
Our aim here is to locate the hard X-ray source in the infrared image
with updated astrometry. The accuracy in absolute astrometry of the
standard processing of the Chandra data (after 2007 May) can be as
good as 0.22 arcsec at the 68\% limit (0.42 arcsec at the 90\% limit)
for bright sources observed near the aimpoint of the ACIS-S. However,
the standard processing of a HST image may have a positional error up
to 1 arcsec. Given the small projected separation ($\approx 1$ arcsec)
of the two nuclei, it is crucial to register the infrared and X-ray
images as accurately as possible.

The HST-ACS I band image (Fig. 1a) is used as a reference for
registering the near IR and the X-ray images as follows. The absolute
astrometry of the ACS image has been corrected using field stars in
the 2MASS catalog, which is only possible with the large field of view
of the ACS and results in the astrometric accuracy to $\sim 0.2$
arcsec. The nuclear region of Mrk 273 in the I band image is confused
by extinction of the complex dust structures and the two nuclei are
not clearly identified. For this reason, we use compact optical
sources which are located outside the dusty nuclear region for
aligning the NICMOS and Chandra images.

Three candidate star clusters near the nuclear regions, seen in both I
and H bands, were used to align the NICMOS image (Fig. 1b) to the ACS
image (Fig. 1a). These objects are labeled in Fig. 1a, and they may
also be identified by their coordinates as GOALS J134442.17+555312.9
(CSC1), J134442.11+555314.5 (CSC2), and J134442.15+555312.5
(CSC3). Since two of the star clusters are faint, the relative
accuracy of the alignment is estimated to be about their beam size
$\sim 0.15$ arcsec\footnote{In the astrometry corrected NICMOS image,
  the 2MASS source is located between the two near IR nuclei, which is
  likely an artifact of the low resolution of the 2MASS survey.}.

The Chandra images were aligned using the bright X-ray source Mrk 273X
(see Fig. 1e), a Seyfert 2 galaxy at $z=0.458$ (Xia et al 1998), which
is listed in the Chandra Multiwavelength Project catalog as CXOMP
J134447.4+555411 (Kim et al 2007) and its optical counterpart is
marked in Fig. 1d. The X-ray source is point-like and has $\approx
1000$ counts in the 0.4-5 keV band, with which the X-ray centroid is
well determined by the detection algorithm {\tt celldetect}. The X-ray
position was displaced by 0.1 arcsec to the east from the optical
nucleus of the galaxy, and a correction has been applied to the X-ray
astrometry. The centroid position of the 4-6 keV source is then found
to be (RA, Dec)$_{\rm J2000}$ = (13h44m42.06s, +55d53m12.77s).

After these registrations, the accuracy of the relative alignment
between the NICMOS and Chandra images is expected to be $<0.2$
arcsec. Kim et al (2007) derived empirical formulae for positional
uncertainty for a Chandra source as a function of source brightness
and off-axis angle of the source location, using image simulations. As
this uncertainty is derived from displacements between the input and
simulated image positions, it can be considered as a statistical
error. The hard X-ray source has $362\pm 9$ counts in the 4-6 keV
band. The uncertainty estimated from the relevant formula is 0.25
arcsec ($1\sigma $). We thus adopt 0.35 arcsec as the absolute
astrometric error. The position of the hard X-ray source and the error
circle are indicated in the NICMOS image (Fig. 1b), which shows that
the hard X-ray source is likely associated with the red, near-infrared
SW nucleus (Fig. 1c).

\subsection{The SE radio component}

Fig. 1c shows an overlay of the VLA 3.6 cm image contours (Condon et
al 1991) on the composite ACS/NICMOS image. No astrometry correction
has been applied to the radio image. There is a slight offset of the
radio image by $\sim 0.15$ arcsec to the south, which appears to be a
remaining alignment error, as the SW radio component has a similar
shift relative to the near-infrared nucleus. As shown in Fig. 1c, the
SE radio component is likely associated with a star cluster seen in
the ACS and NICMOS images, as first pointed out by Scoville et al
(2000). High resolution radio imaging shows that this source is
extended and has a steep spectrum (Carilli \& Taylor 2000; Bondi et al
2005). These radio observations suggest a starburst as its origin
(Bondi et al 2005) but they pointed out that the lack of a near-IR
counterpart was a problem. An association of the radio source with the
star cluster seen in optical and near infrared would favour a
starburst interpretation (further discussed in \S 4.1).

\subsection{Possible extension of 6-7 keV emission}

The 4-6 keV emission, used in the imaging analysis above, is dominated
by an absorbed continuum of the active nucleus with negligible
contribution from extended emission seen at lower energies. Fig. 2a
shows the radial profiles of the 4-6 keV emission in the northeast
(${\sl PA}=45^{\circ}$) and southwest (${\sl PA}=225^{\circ}$) halves,
indicating the 4-6 keV source is approximately symmetric in the
northeast-southwest direction and close to the point spread function
(PSF). However, in the 6-7 keV band (Fig. 2b), the northeast half shows a
$\sim 3\sigma$ excess over the southwest half, which is consistent with the
PSF.
This excess carries about 10\% of the total 6-7 keV emission (121 counts)
and is found around the N nucleus, as shown in the (6-7 keV)/(4-6 keV)
ratio image (Fig. 2c). 

The 6-7 keV band contains the 6.4 keV Fe K line (at 6.17 keV in the
observed frame). The spectrum taken from a $1^{\prime\prime}$
  diameter aperture of the 6-7 keV excess region, as indicated in
  Fig. 1e (inset), indeed shows a strong 6.4 keV Fe K line with
  equivalent width, EW $=1.0\pm 0.5$ keV (labeled as 'N' in
  Fig. 3). For a comparison, another spectrum was taken from a region
  centred on the SW nucleus with a $1.5^{\prime\prime}$ diameter
  aperture (as indicated in Fig. 1e--inset), which does not overlap
  with the above 6-7 keV excess region. This spectrum (labeled as 'SW'
  in Fig. 3) is essentially of the absorbed active nucleus and the 6.4
  keV Fe K line has EW $=0.23^{+0.09}_{-0.07}$ keV, in agreement with
  that measured for the total emission of the objects (Ptak et al
  2003; Xia et al 2002; Iwasawa et al 2011). Thus the Fe K line in the
  N spectrum is much stronger (with respect to the contiuum) than in
  the SW spectrum.

This is consistent that the 6-7 keV excess is mainly due to an
enhancement of the Fe K line, suggesting possible presence of an extra
source of Fe K, albeit the evidence is marginal given the limited
statistics. If it is real, a reflection-dominated spectrum, as seen in
a heavily obscured AGN, is a good candidate for this extra source,
since it contributes primarily to the Fe K line but significantly less
to the continuum. The 3-8 keV continuum of the N spectrum is flat
($\Gamma = 0.0^{+0.4}_{-0.6}$) and shows no obvious sign of an
absorption cut-off, as seen in the SW spectrum. This is also
consistent with a reflection spectrum but the absorption cut-off in
the lower energies may be masked by strong soft X-ray emission.  The
reality of the image extension and the Fe line enhancement there need
to be investigated further with higher quality (both in spatial
resolution and counting statistics) data. The Fe K excess region
spatially coincides with the N nucleus, which could lead to an
interesting possibility, as discusse below (\S 4.1).

This region is also brightest in soft X-rays, particularly in the 1-3
keV band (Fig. 1e--inset). The soft X-ray spectrum shows
emission-lines of H-like Ne, Mg and Si that are comparable or even
stronger than their He-like counterparts (as marked in Fig. 3),
indicating high-ionization. If these lines are of thermal origin, the
temperature should be $kT \simeq 1.5\pm 0.3$ keV, as inferred by a
thermal spectrum fit to the energy range of the Mg and Si lines. This
temperature is much higher than the mean temperature of the bright
central part, e.g., inner 10$^{\prime\prime}$ area ($kT\sim 0.8$ keV,
Xia et al 2002; Grimes et al 2005). Note that Ne {\sc ix} (0.9 keV) is
not expected at $kT\sim 1.5$ keV, and its detection perhaps
suggests the presence of multi-temperature gas in this region.


\begin{figure}
\centering
\centerline{\includegraphics[width=0.5\textwidth,angle=0]{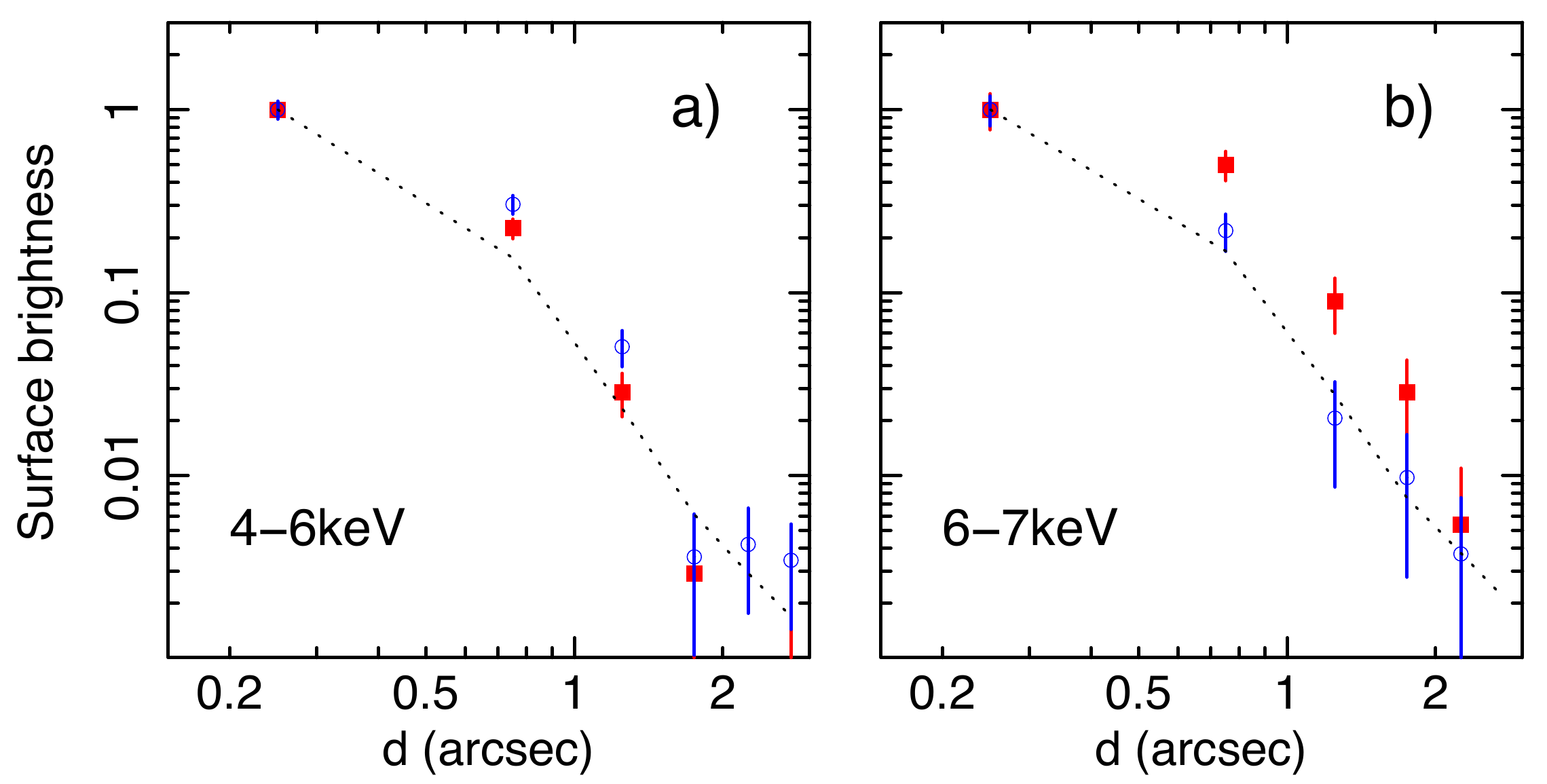}}
\rightline{\includegraphics[width=0.25\textwidth,angle=0]{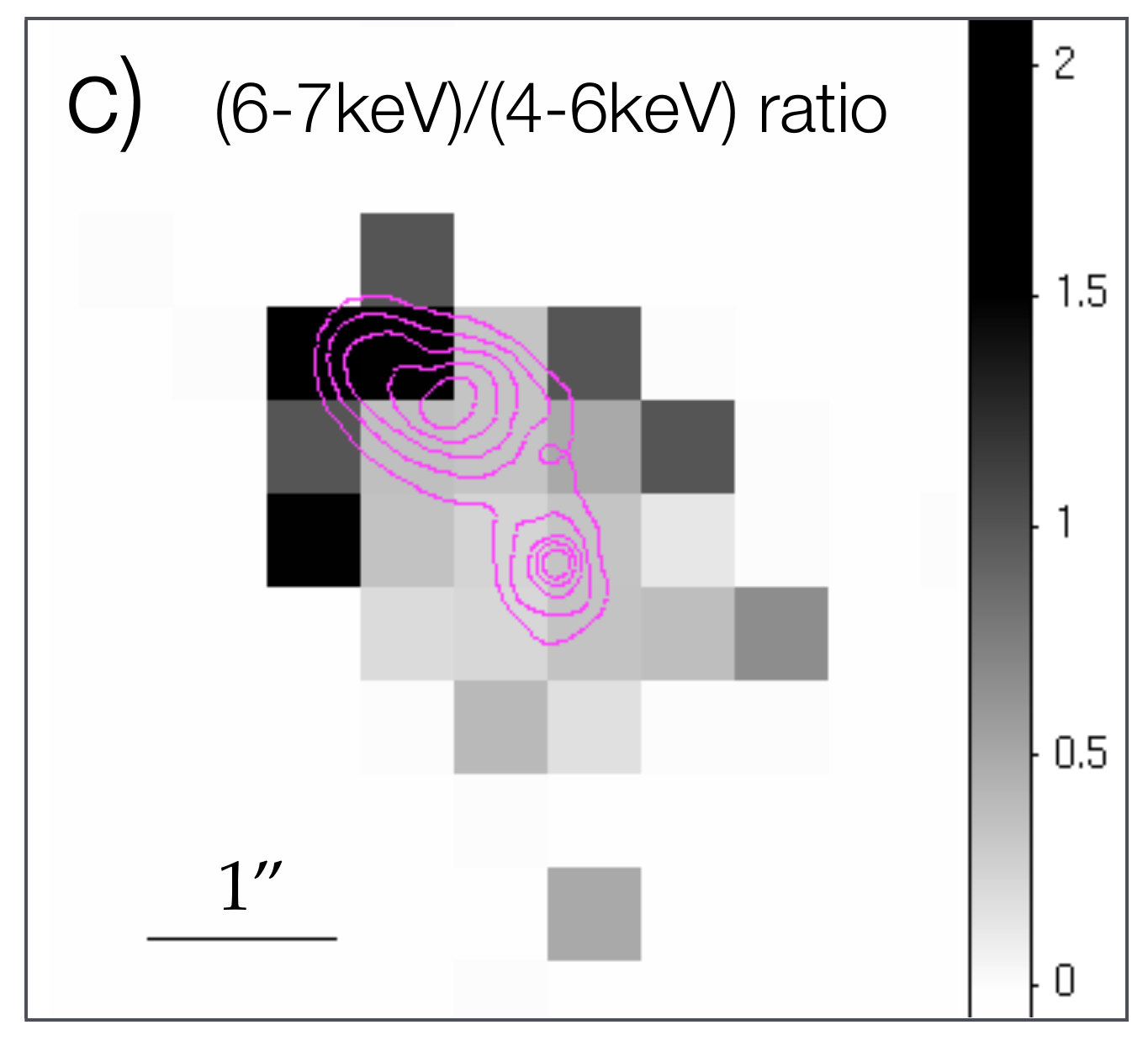}}
\caption{a) The radial surface brightness profiles of the NE (red
  squares) and SW (blue circles) halves in the 4-6 keV band, with the
  simulated PSF at 4.51 keV (dotted line); b) Same as a) but in the
  6-7 keV band, with the similated PSF at 6.4 keV (dotted line); c)
  The (6-7 keV)/(4-6 keV) ratio image of the nuclear region of Mrk
  273, overlaid by the NICMOS H-band image contours showing the
  N and SW nuclei. The contours are drawn at five logarithmic
  intervals with the lowest level at 20\% intensity of the peak
  brightness at the centre of the SW nucleus. The orientation of the
  image is north up, east to the left. The scale bar indicates 1
  arcsec. This map shows a trend of a 6-7 keV band enhancement towards
  NE. }
\end{figure}


\begin{figure}
\centerline{\includegraphics[width=0.4\textwidth,angle=0]{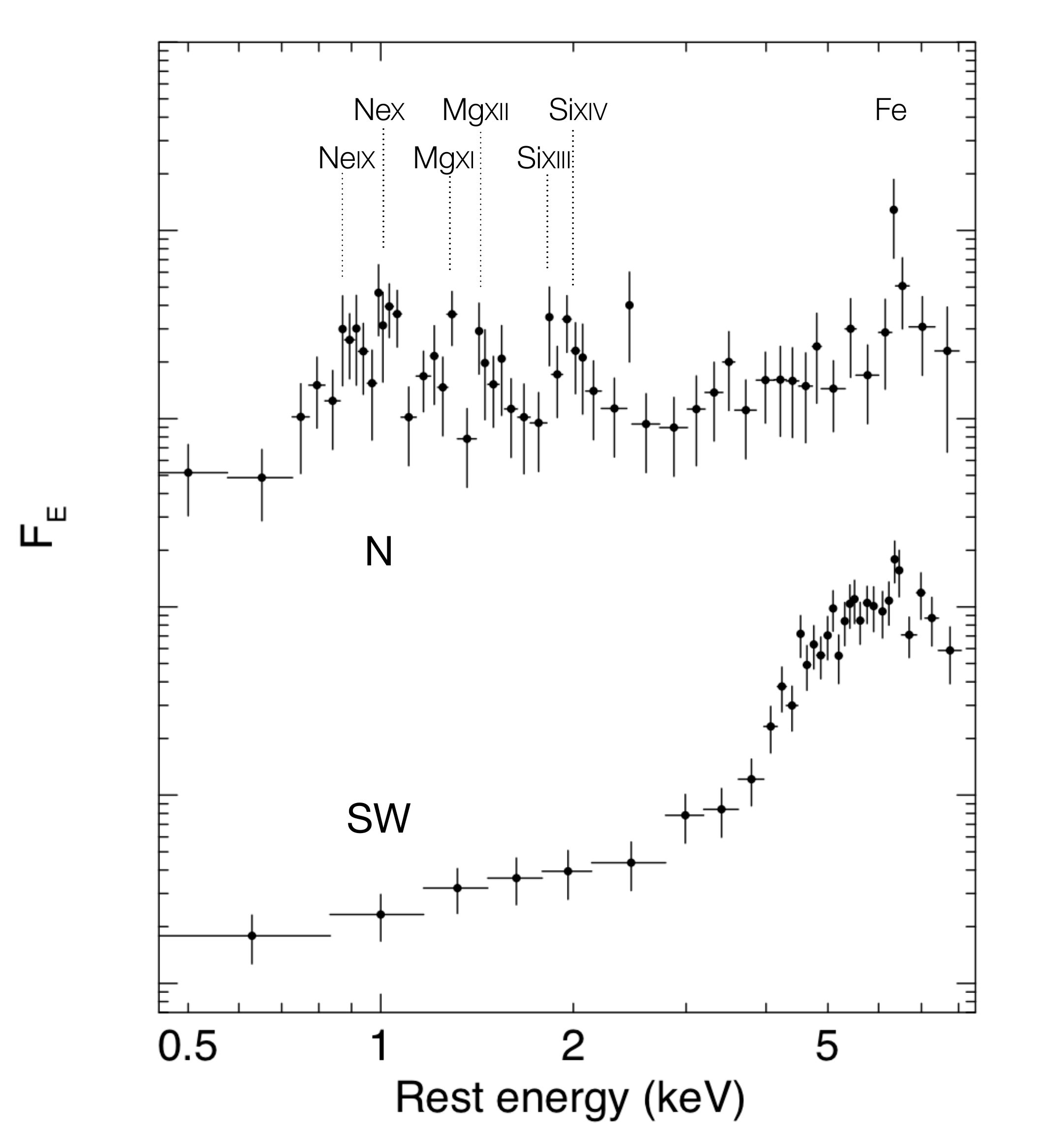}}
\caption{The Chandra ACIS-S spectra of the 6-7 keV excess region
  (Fig. 2c) at the N nucleus and of the SW nucleus. The spectral
  apertures are 1 arcsec and 0.75 arcsec in radius, respectively, and
  are shown in Fig. 1e (inset). The spectra are presented in flux
  density unit and arbitrarily shifted for visual clarity, and the
  energy scale has been corrected for the galaxy redshift. The N
  spectrum shows a possibly larger EW of Fe K at 6.4 keV than the SW
  spectrum and the soft X-ray lines indicating high ionization (e.g.,
  H-like Mg and Si), as marked.}
\end{figure}

\subsection{The extended soft X-ray nebula}

\subsubsection{A shadow of the tidal tail}

\begin{figure*}
\centerline{\includegraphics[width=0.7\textwidth,angle=0]{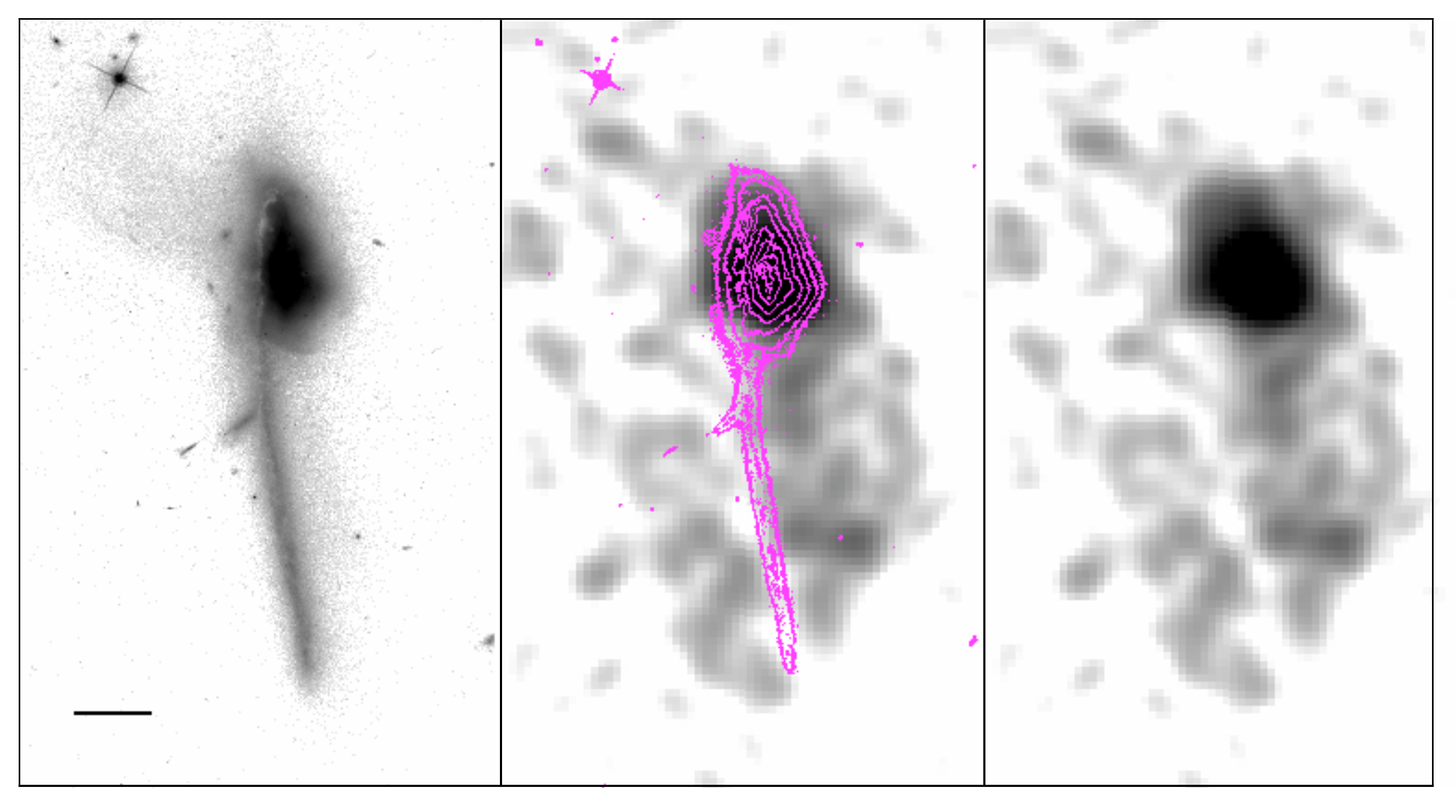}}
\caption{Left: The HST-ACS {\sl I}-band (F814W) image of Mrk 273;
  Middle: The 0.4-1.1 keV Chandra image of Mrk273, overlaid by the
  contours obtained from the optical {\sl I}-band image; Right: The
  same 0.4-1.1 keV Chandra image as in the middle panel. The X-ray
  image of 1 arcsec resolution has been smoothed using a circular
  Gaussian kernel with the dispersion of 1.8 arcsec. The contours for
  the HST optical image are drawn at 10 logarithmic intervals. In the
  right panel, a dark lane runs through the soft X-ray nebula in the
  N-S direction, which coincides with the area where the optical tidal
  tail is preent. The orientation of the images is north up, east to
  the left. The scale bar corresponds to 10 arcsec.}
\end{figure*}

A large soft X-ray nebula extending to the south is clearly visible in
the Chandra image (Fig. 1e), as noted by several authors (Xia et al
2002; Gonz\'alez-Mart\'in et al 2006; Ptak et al 2003; Grimes et al
2005). Comparison of the soft X-ray and optical images shows that the
long, optical tidal tail extending to the south is remarkably well
aligned with a dark gap running through the soft X-ray nebula in the
N-S direction (Fig. 4). The position angle (PA) of the X-ray dark lane
is $\sim 10^{\circ}$, which increases to $\sim 15^{\circ}$ at the
southern end while the optical tidal tail has PA$\sim
10^{\circ}$. This can be understood if the tidal tail is located in
front of the soft X-ray nebula in our line of sight such that soft X-ray
emission is absorbed by the cold gas in the tidal tail.

The ACIS-S spectrum of the southern X-ray nebula (Fig. 5), taken from
the region indeicated in Fig. 1e, shows that the bulk of the light is
emitted below 1.1 keV. The total source counts in the 0.4-1.1 keV band
are $\sim 300$ counts. To make emission of such a spectrum totally
suppressed
at energies below 1 keV, an absorbing column density of \nH $\approx
1\times 10^{22}$ \psqcm\ or larger is required. However, as the X-ray
source is diffuse, the limiting column density would be slightly
relaxed. The projected area of the tidal tail's shadow is measured to
be $\approx 2.5\times 38$ arcsec$^2$. Under a crude assumption of a
uniform brightness distribution over the nebula, we adopt detection of
$2\sigma $ counts in more than half the area is the limiting
condition. Assuming the spectral model for the nebula (\S 3.4.2),
expected counts from the shadowed area are simulated for incremental
$N_{\rm H}$ values of absorbing matter at the galaxy redshit. As
$N_{\rm H}$ is increased by an interval of $\Delta N_{\rm H} = 1\times
10^{21}$ cm$^2$, expected counts go below the detection limit, when
$N_{\rm H}$ exceeds $6\times 10^{21}$ cm$^2$. We consider this $N_{\rm
  H}$ as the lower limit of the column density of cold gas in the
tidal tail, which makes the shadow on the soft X-ray nebula. This
column density is similar to that of an edge-on galaxy disk. Similar
measurements of disk shadowing have been made against the
extragalactic X-ray background (e.g., Barber, Roberts \& Warwick
1996).


\subsubsection{Spectral properties}


\begin{figure}
\centering
\centerline{\includegraphics[width=0.35\textwidth,angle=0]{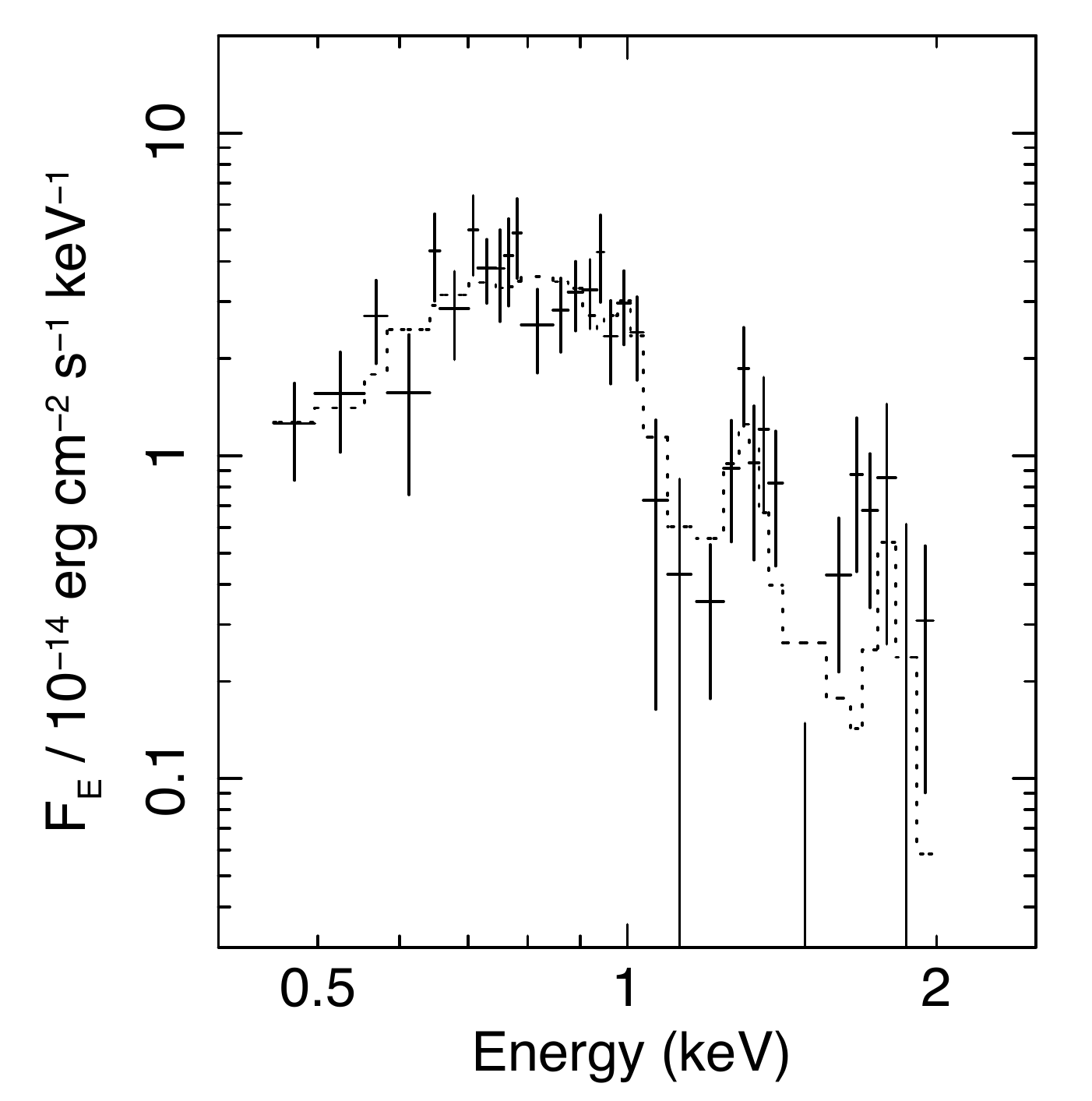}}
\caption{The ACIS-S spectrum of the southern X-ray nebula of Mrk
  273. The dotted-line histogram shows the best-fit thermal emission
  spectrum (see text for details). Note strong emission-lines from Mg
  {\sc xi} and Si {\sc xiii} in the 1-2 keV range.}
\end{figure}

While the spectrum of the nebula (Fig. 5) is dominated by the soft
X-ray emission below 1 keV, there is weak but significant emission up
to 2 keV containing $\sim 50\pm 10$ counts, mainly due to line
emission of Mg {\sc xi} at 1.34 keV and Si {\sc xiii} at 1.85
keV. There is also an excess on the lower energy side of Si {\sc
  xiii}. These strong lines suggest an enhancement of metallicity in
$\alpha $ elements. Fitting the spectrum with a thermal emission model
computed by MEKAL (Kaastra 1992), in which the abundances of elements
relavant to the 0.4-2 keV band are left variable, gives a metallicity
of {\sl Z}(Ne, Mg, Si) $=(3\pm 1) Z_{\odot}$, {\sl Z}(Fe)$=(0.4\pm
0.3) Z_{\odot}$, and {\sl Z}(O)$= (0.7\pm 1) Z_{\odot}$ with a
temperature of $kT = 0.47^{+0.08}_{-0.05}$ keV and the Galactic
absorption $N_{\rm H}=9\times 10^{19}$ cm$^{-2}$ (Kalberla et al
2005). The relatively small Fe and O abundances reflect weakness of
the Fe L emission bump around 1 keV and OVIII at 0.65 keV. This result
is in agreement with Grimes et al (2005), but differs from the low
metallicity obtained by Xia et al (2002) who used data only below 1
keV, where the Mg and Si lines are not present, and with the solar
abundance pattern.

Enhancement of metallicity in $\alpha $ elements is expected in the
interstellar medium enriched by ejecta of core-collapse supernovae,
which a starburst would mainly produce. The nebular spectrum is
similar to that found in the extended nebula of another ULIRG Mrk 231
in terms of temperature and the strong Mg and Si lines (Iwasawa et al
2011). The apparently lower metallicity of O than the other $\alpha
$-elements is not compatible with the classical theoretical prediction
(e.g., Nomoto et al 1997) but similar to that found in well studied
starbursts in NGC 6240 (Netzer et al 2005) and M 82 (Ranalli et al
2008).

\section{Discussion}

\subsection{Nature of the double nucleus}

With the registration with the HST image, the SW nucleus appears to be
the location of the hard X-ray source in Mrk 273, which is absorbed by
cold gas of \nH $\sim 4\times 10^{23}$ \psqcm.  This is in agreement
with the suggestion by Scoville et al (2000) that the SW nucleus, a
red, unresolved NICMOS source, is the location of an active nucleus
and other properties described in \S1. Besides the Seyfert 2 spectrum
at the SW nucleus, Colina et al (1999) also pointed out that the
kinematically quiescent [O{\sc iii}] component, located $\sim
2^{\prime\prime}$ further SW of the SW nucleus, is similar to the
ionization-cones around Seyfert 2 galaxies. The mid-IR AGN indicator
[Ne {\sc v}] may therefore originate from this region. The
mid-infrared to hard X-ray ratio, log $L_{12\mu {\rm m}}/L_{\rm
  6keV}\simeq 1.2$, of the SW nucleus is comparable to those normally
observed in AGN (e.g., Elvis et al 1994; Lusso et al 2010). While the
radio emission is weak, the ratio of flux densities $f_{\nu}(12.5\mu
m)/f_{\nu }(8.4 GHz) = 290$, as derived by Soifer et al (2000), is not
unusual for a radio-quiet AGN.

The far-infrared portion of the SEDs of the two nuclei are not
known. The radio band is dominated by the N component, with log
$P_{\rm 1.4GHz}\simeq 27.7$ W Hz$^{-1}$. If the far infrared emission,
with log $P_{\rm 70\mu m}\simeq 29.9$ W Hz$^{-1}$ (estimated by an
interpolation of the IRAS 60 $\mu $m and 100 $\mu $m fluxes of Sanders
et al 2003), is also dominated by the N component, then the 1.4 GHz
and 70 $\mu $m luminosities are consistent with the well-known
far-infrared/radio correlation (e.g., Appleton et al 2004). This
suggests that the N nucleus is likely powered by a starburst and the
major source of the far-infrared emisison where the bolometric power
peaks ($\sim 60\mu$m). A similar conclusion can be reached by an
argument based on the AGN luminosity of the SW nucleus. The
absorption-corrected 2-10 keV luminosity of the hard X-ray source is
$0.9\times 10^{43}$ \ergps. Assuming the bolometric correction of
Marconi et al (2004), the bolometric luminosity of the AGN in the SW
nucleus is estimated to be $2\times 10^{44}$ \ergps or log $L^{\rm
  bol}_{\rm AGN}/L_{\odot}\sim 10.7$, which is a factor of $\sim 30$
smaller than the total $L_{\rm IR}$. Relative to this X-ray estimate,
the mid-IR diagnostics with [Ne V] $\lambda 14.3\thinspace\mu $m, [O
IV] $\lambda 25.9\thinspace\mu $m and PAH emission at 6.2 $\mu $m seem
to suggest a larger (but not a dominant) contribution from AGN (Armus
et al 2007; Petric et al 2010).

Downes \& Solomon (1998) suggested that the N nucleus is an extreme
compact starburst with a high luminosity-density, similar to the
western nucleus of Arp 220. High-resolution radio continuum imaging
supports this hypothesis (Carilli \& Taylor 2000; Bondi et al 2005).
The LINER-type excitation and the adjascent, kinematically turbulent
[O{\sc iii}] component (Colina et al 1999) can be explained by a
strong outflow from the starburst. As discussed above, the total
far-infrared output of Mrk 273 is likely dominated by a starburst
occuring in this region. As shown in Fig. 1a, bright soft X-ray
emission is observed here. The soft X-ray spectrum shows thermal
characteristics and high metallicity (Xia et al 2002; Ptak et al 2003;
Grimes et al 2005). 

At energies above 4 keV, any contribution of the N nucleus appears to
be negligible. A possible exception is in the 6-7 keV range. If the
image extension in this energy range is real and due to enhanced Fe K
emission, as discussed in \S 3.3, this presents a possibility that a
heavily absorbed (possibly Compton-thick) AGN may exist in the N
nucleus as well. Assuming a reflection-dominated spectrum with a 6.4
keV Fe K line with EW = 1 keV, an expected 4-7 keV flux of this
(hypothetical) source would be $\sim 2\times 10^{-14}$ \ergpspsqcm,
which is only 5\% of the total observed flux from the whole system in
the same band. Even if an intrinsic source flux is 2 orders of
magnitude larger, its 2-10 keV luminosity would be about $10^{43}$
\ergps, similar to that of the SW and, in any case, have a minor
contribution to the bolometric luminosity. It should be noted that the
6-7 keV excess could also be due to extended Fe K emission induced by
AGN in the SW nucleus, as seen in NGC 4388 (Iwasawa et al 2003),
rather than originating from the N nucleus. At the current data
quality, evidence for an AGN in the N nucleus is thus weak. However,
this raises an exciting possibility that Mrk 273 is another dual AGN
system similar to the cases of NGC 6240 (Komossa et al 2003) and Mrk
266 (NGC 5256, Mazzarella et al 2011).

\subsection{Cold gas in the tidal tail shadowing soft X-ray nebula}

A rough estimate of the mass of cold gas in the tidal tail responsible
for shadowing the soft X-rays can be obtained from the minimum column
density ($N^{\rm lim}_{\rm H}=6\times 10^{21}$ cm$^{-2}$) which is
required to absorb away the soft X-ray photons from the nebula (\S
3.4.1). We note that a dust lane extends across the tidal tail from
the main body of the galaxy, as seen in the HST optical image
(Fig. 1d). If cold gas is distributed uniformly in a cylinder with a
diameter of 2.5 arcsec ($d\sim 2$ kpc) and a length of 40 arcsec
($l\sim 30$ kpc) lying on the plane of the sky, the mean density would
be $n=N^{\rm lim}_{\rm H}/d\sim 1$cm$^{-3}$. Thus the gas mass is
estimated to be $\geq 2\times 10^9 M_{\odot}$. However, a more likely
geometry of this tidal tail is an edge-on disk, as argued in \S 1 and
\S 3.4. In this case, the average density would be lower.

Similarity in morphology between the soft X-ray and H$\alpha $
emission has been noted by Xia et al (2002) and Grimes et al
(2005). This correspondence is often found in objects with
galactic-scale outflows, such as Arp 220 and NGC 6240 (e.g., Veilleux,
Cecil \& Bland-Hawthorn 2005; Armus et al 1990). The enhancement of
$\alpha $-elements in the soft X-ray emitting gas (\S 3.3.2) supports
prior interpretation of the diffuse nebula as a starburst-driven
superwind. Although the outflow velocity at inner radii is as large as
2400 km s$^{-1}$ (Colina et al 1999), it will slow down quickly until
escaping from the galaxy's potential. To reach 30 kpc, which is
roughly the projected length of the southern extension of the nebula,
the required time is $t\sim 10^8 (v_{300})^{-1}$ yr, where $v_{300}$
is the average outflow velocity in unit of 300 km s$^{-1}$. This can
be a crude measure of the starburst age. The linearity, thinness and
high surface-brightness suggests that a tidal tail (remnants of a
galaxy disk) is being viewed nearly edge-on, trailing a northward
orbital encounter. Assuming the component of the relative orbital
velocity in the plane of the sky is $\sim 300$ km s$^{-1}$, material
in the tidal tail $\sim 30$ kpc south of the nuclei would have been
drawn out from the parent galaxy disk $\sim 10^8$ yr ago. If the
starburst was episodic or roughly continuos over the duration of the
ongoing merger, as suggested by simulations, the alignment of the
diffuse X-ray emitting nebula and the stellar tidal tail is not a
coincidence, but rather an artifact of the main power source of the
starburst-driven wind moving northward during the creation of the
trailing tidal tail.

\section{Summary}

\begin{itemize}
\item With the new registration of the Chandra X-ray image with the optical
and near-IR images from HST, we identified the absorbed, hard X-ray
source of the Seyfert 2 nucleus in Mrk 273 with the SW nucleus, which
revises the previous identification to the N nucleus.

\item The hard X-ray source is point-like except in the 6-7 keV band,
  where the image shows a slight extension to the northeast. This
  excess appears to be due to enhanced Fe K line emission at 6.4 keV
  around the N nucleus. A possible implication is that a heavily
  obscured active nucleus might be present also in the N nucleus,
  while a starburst remains to be a dominant source of the far-IR
  luminosity.

\item A dark lane in the soft X-ray nebula extending to the south is found
to align well with the optical tidal tail. This probably means the
tidal tail is located in front of the soft X-ray nebula and absorbs
the X-ray photons along our line of sight. The column density of cold gas
in the tidal tail is estimated to be $N_{\rm H}\geq 6\times 10^{21}$
cm$^{-2}$.

\end{itemize}

\begin{acknowledgements}
  This research made use of archival data maintained at the Chandra
  X-ray Center (CXC) and Space Telescope Science Institute (STScI),
  data products from the Two Micron All Sky Survey (2MASS), and the
  NASA/IPAC Extragalactic Database (NED).
\end{acknowledgements}

\end{document}